\documentclass{camera}
\usepackage{graphicx}  % uncomment this if your want to insert figures

\begin{document}

%%%%%%%%%%%%%%%%%%%%%%%%%%%%%%%%%%%%%%%%%%%%%%%%%%%%%%%%
% The title, only the first letter capitalized; if you want to split it in
% two or more lines, put a \\ macro at each line break
% example: 
%   \title{Title: first line\\ second line}
%
\title{Elie:\\ an event generator for nuclear reactions}

%%%%%%%%%%%%%%%%%%%%%%%%%%%%%%%%%%%%%%%%%%%%%%%%%%%%%%%%
% The author(s), separated by commas; do not put a
% comma before the last author, use instead the \and
% macro which produces a normal ``and'' in the
% caps/small caps context
%
\author{Dominique Durand}

%%%%%%%%%%%%%%%%%%%%%%%%%%%%%%%%%%%%%%%%%%%%%%%%%%%%%%%%
%
\organization{LPC Caen-ENSICAEN-Universit\'e de Caen-Basse-Normandie\\
durand@lpccaen.in2p3.fr}

\maketitle

\begin{abstract}
An event generator for the description of nuclear reactions in the Fermi
energy range is briefly introduced and first comparisons with experimental data 
are shown.
\end{abstract}

%%%%%%%%%%%%%%%%%%%%%%%%%%%%%%%%%%%%%%%%%%%%%%%%%%%%%%%%
% Write the text starting from here and using the usual
% LaTeX commands.
%
\section{Introduction} 

For several decades now, nuclear collisions in the Fermi energy range have been used to 
explore the fundamental 
properties of nuclei under extreme conditions of pressure, temperature and/or angular momenta \cite{livre}. 
From an experimental point of view, large detection facilities-the so-called $4\pi$ detectors-
have been developped to detect most of the emitted charged (sometimes also neutral) particles \cite{general}. 
Hence, complex events are recorded with high multiplicities. This leads to rather sophisticated 
analyses which 
may be affected by kinematical cuts and/or by uncontrolled selection criteria. Thus, a safe 
comparison between theory and
experiment requires simulation tools based on models as realistic as possible and 
based on assumptions that can
be safely tested. There are essentially two modes of description of the data.

\begin{itemize}
\item
the microscopic transport models which in
principle, should be the best solution. However, calculations are 
usually lengthy and the contact with experimental data may in some cases be difficult \cite{general}.
\item
the statistical models based on equilibrium hypotheses (such as SMM for instance) 
are widely used and have met some successes \cite{smm}. 
However, they rely on assumptions which may not be relevant in the context of 
nuclear reactions. More, their comparaison with experimental data is performed 
only after severe (and maybe uncontrolled) event and particle selections.      
\end{itemize}

Our aim in this work 
is to explore within a schematic model what happens
when the hypothesis underlying the equilibrium statistical models are abandoned. 
In particular, we wish to take into account explicitely the entrance
channel characteristics -the geometry- of the reaction as well as the initial correlations 
of the nucleons in momentum space, namely the internal Fermi motion.  

\section{Brief description of the model}

The ELIE event generator \cite{prepare} is based on a two-step scenario of the reaction: 
\begin{itemize}
\item
an entrance channel phase ending with the formation (for finite impact parameters) of a 
projectile-like fragment, 
a target-like fragment (both being moderately excited) and participants whose 
partition (Intermediate Mass Fragments (IMF) and light particles) is
obtained by means of a random process (see later)
\item
a second phase considering secondary decay and propagation towards the detectors.  
\end{itemize}

The geometry of the collision is borrowed from the high energy participant-spectator 
picture: the mass numbers of
the projectile-like, target-like and participants are obtained by considering the geometrical overlap 
of the nuclei for each impact parameter. To build the kinematics of the projectile-like, the target-like 
and the partition of the participants, the following 
hypothesis are assumed:
\begin{enumerate}
\item
the momentum distribution of the incoming
nucleons inside the two partners is supposed to have no time to relax on a time scale comparable with the
reaction time. This is a frozen approximation: only a few hard nucleon-nucleon 
collisions can occur and those latter are governed by a single
parameter: the mean free path.

\item
the partition of the participants is generated by a random process in momentum space. The mass 
number A of each species (including A=1 free nucleons) is sequentially chosen at random by
picking A nucleons from the nucleon momentum distribution. For IMF's ($A \geq 4$),    
the excitation energy, $E^*$, 
is obtained by summing the center-of-mass kinetic energy
of all nucleons belonging to the fragment. If $E^*$ is larger 
than a maximum value associated with a maximum temperature $T_{max}$, the fragment is rejected and a new try is
made untill all nucleons have been assigned. In the following, it turns out 
that a value of $T_{max}$ = 5.5 MeV
allows
to reproduce the experimental data (with a level density parameter equal to A/10, $E^*$=3MeV/u). This 
is in agreement 
with causality: the fragment lifetime should 
be at least comparable with the reaction time. In the Fermi 
energy range, this latter is of the order of a few
tens of fm/c, thus leading to a maximum temperature 
around 5 MeV and to excitation energies close to 3-4 MeV/u. 

\item
the N/Z content of the projectile-like is taken equal to the N/Z of the projectile 
and the same rule is applied to the
target-like and to the participants. For light particles ($Z \leq 2$), all existing isotopes are randomly
considered.  

\item
the projection of the partition in real space is accomplished by propagating the
fragments (starting from the origin of space) 
according to their initial
velocity assuming straight line trajectories untill  
there are no more geometrical overlaps of particles.

\end{enumerate}

{\bf In short, the model considers all possible random 
partitions compatible with geometry, 
conservation laws, a maximum internal temperature of about 5.5 MeV, a N/Z 
memory of the entrance channel 
and a nucleon momentum distribution close to the
initial one.} 
\\

In a second step, the partition is  
propagated in space-time and secondary decays are considered.
This is done using the
SIMON event generator \cite{simon}. Two major advantages of our approach are
that 
it considers all impact 
parameters and, for each event,
considers all nucleons. In particular, there is no distinction between 
equilibrium and pre-equilibrium particles. 
As such, there is  no need for selection criteria and/or kinematical 
cuts and a full comparison with the
experimental data is made possible.

\section{Comparisons with experimental INDRA data}
INDRA data \cite{ref-xesn} has been selected requiring 
that at least 80 $\%$ of the total charge and total linear momentum for 
charged particles emitted in the forward
direction in the centre-of-mass  
be detected. A tensor based on the momenta of the fragments is built and 
its diagonalisation 
gives three eigen-values with which  
the so-called $\theta_{flow}$ 
angle is built. This is the angle between the main axis of the tensor and the beam axis.  
We consider only events with $\theta_{flow}$ larger than 25 degrees.
In our model, 
this angular range corresponds to reduced impact parameter 
lower than .5 and centered 
around .3 (central collisions). 
\begin{figure}[h]
\begin{center}
\includegraphics{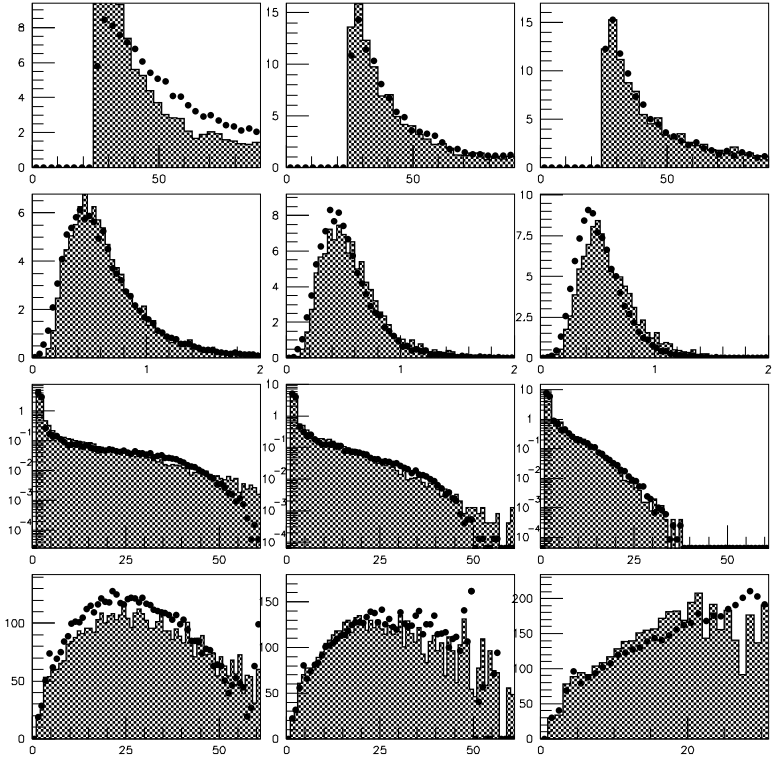}
\caption{Comparison between INDRA data (black points) and ELIE data (histograms) for central Xe+Sn 
collisions at 25 (left), 32 (middle) and 50 MeV/u (right). From up to down:
flow angle (in degrees) distribution, isotropy ratio, charge ($Z$) distribution, mean kinetic energy (in MeV) 
as a function of $Z$.}
\label{fig01} % optional figure label, must be unique
\end{center}
\end{figure} 
The flow angle distribution as well as the isotropy ratio are correctly reproduced by the model (see Figure
\ref{fig01}). This latter requires a mean free
path evolution from 30 fm at 25 MeV/u down to 10 fm at 50 MeV/u. The agreement for charge distributions 
and kinetic energies shows that a random process and an account of the nucleonic 
Fermi motion allow to reproduce experimental data without 
invoking a compression/expansion scenario as often assumed in
multifragmentation statistical models. We show some results about light 
charged particles in Figure \ref{fig02}. 
\begin{figure}[h]
\begin{center}
\includegraphics{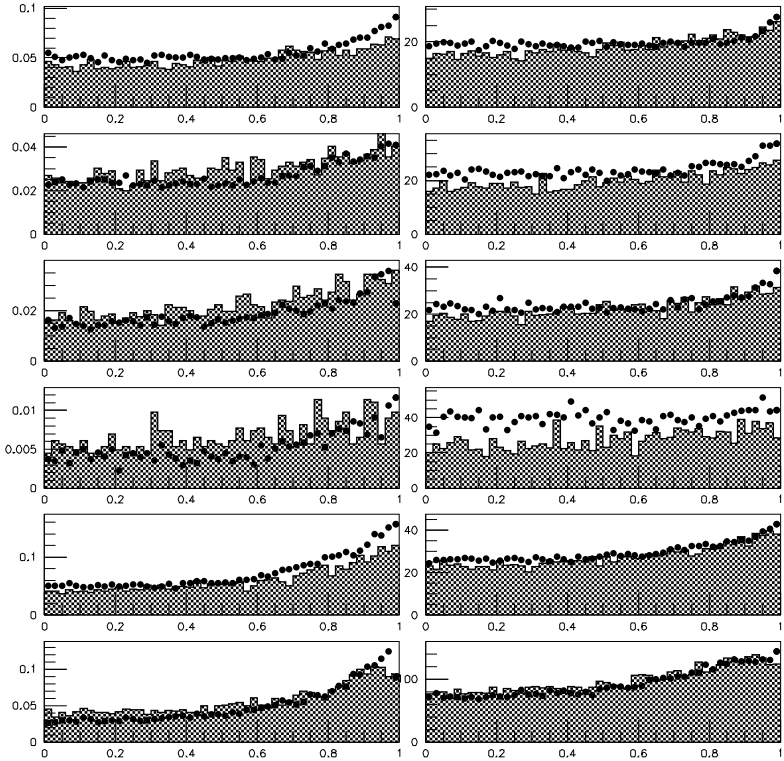}
\caption{Comparison between INDRA data (black points) and ELIE data (histograms) for central Xe+Sn 
collisions at 32 MeV/u. Left: mean multiplicity as a function of the c.o.m cosine (limited to the forward
direction), right, mean kinetic energy (in
MeV)
as a function of the c.o.m cosine. From up to down: protons, deuterons, tritons, He3, alphas, IMF's.}
\label{fig02} % optional figure label, must be unique
\end{center}
\end{figure} 
We recall that most 
of them are produced by picking
randomly nucleons from the initial two Fermi spheres. However, some of them are 
evaporated by the excited fragments on longer time scales.  
The mean number and kinetic energy
as a function of the CM angle as predicted by ELIE are in correct agreement with INDRA data. 
We now consider a comparison of the model with minimum bias data for Ni+Ni reactions at 82
MeV/u (Figure \ref{fig03}) \cite{ref-nini}. This corresponds to collisions for which only the
completeness criterium for the forward-emitted particles has been considered.
The optimum value for the mfp at such an energy 
is found close to 5 fm. Thus, the mfp evolves from 30 fm at 25 MeV/u down to 5
fm at 82 MeV/u. This decrease is interpreted as the opening of 
the phase space for in medium 
nucleon-nucleon collisions due to a reduction of the Pauli blocking factor. 
An overall good 
agreement is obtained although the model overestimates the A=3 species. Notice that the rise and fall of the
fragment multiplicity as a function of the total multiplicity is correctly reproduced. The kinematical
obervables (not shown here) are also in good agreement.                  
\begin{figure}
\begin{center}
\includegraphics{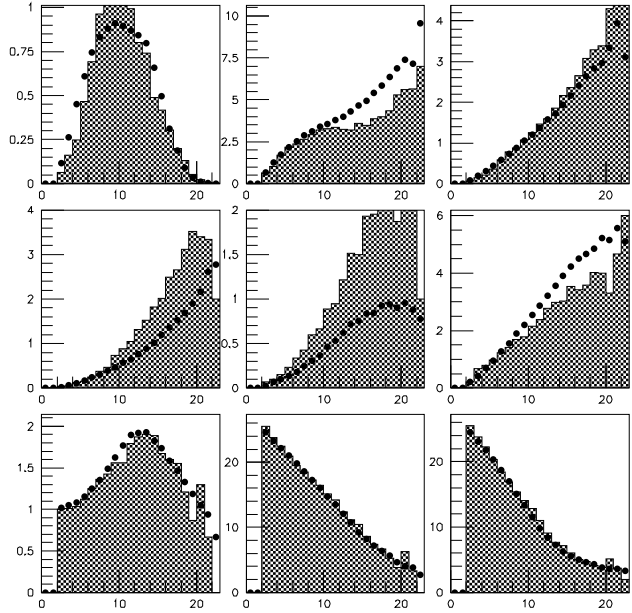}
\caption{Comparison between INDRA data (black points) and ELIE data (histograms) for minimum bias Ni+Ni 
collisions at 82 Mev/u. Left,up; total multiplicity $M_{tot}$. Then, from left to right and from up to low,
proton, deuteron, triton, He3, alpha and IMF mean 
multiplicity as a function of $M_{tot}$, then $Z_{bound}$ (total charge of IMF'S) and $Z_{max}$ 
as a function of
$M_{tot}$}
\label{fig03} % optional figure label, must be unique
\end{center}
\end{figure} 
\section{Summary}
Simulated data produced by the ELIE event generator have been compared 
with INDRA data.  
The kinetic energy and angular 
distributions of both fragments and light charged particles are well reproduced. 
Most light particles 
as well as fragments are produced rapidly at nearly normal density by a random  
process. This latter is constrained by geometry, conservation laws and causality which is 
expressed by the fact that fragment lifetime
should be longer than the reaction time. 
Thus, the projectile-like, the target-like and the fragments from the participant zone 
emerge from the reaction at a moderate excitation energy close to 3 MeV/u 
corresponding to a maximum temperature of about 5.5 MeV in agreement with internal temperature cluster
measurements whatever the incident energy and the system considered. 
The strong memory of the entrance channel (transparency) as well as the internal motion of the 
nucleons inside the two partners 
of the reaction play a crucial role in our approach and are necessary to reproduce the experimental data. 
We thus believe that the present model is a valuable alternative to thermal 
statistical approaches based on equilibrium at low density.

% For Figures insertion uncomment the next section

%\begin{figure}
%\includegraphics{figurename}
%\caption{Your caption here}
%\label{fig01} % optional figure label, must be unique
%\end{figure}

%%%%%%%%%%%%%%%%%%%%%%%%%%%%%%%%%%%%%%%%%%%%%%%%%%%%%%%%
% End of the paper
%
\end{document}